\documentclass[journal,article,submit,pdftex,moreauthors]{Definitions/mdpi} 

\usepackage{placeins}  
\usepackage{subfig}

\firstpage{1} 
\pubvolume{1}
\issuenum{1}
\articlenumber{0}
\pubyear{2023}
\copyrightyear{2023}
\datereceived{ } 
\daterevised{ } 
\dateaccepted{ } 
\datepublished{ } 
\hreflink{https://doi.org/} 

\Title{Phase transition in the Galam's majority-rule model with information-mediated independence}

\TitleCitation{Title}

\Author{Andr\'e L. Oestereich $^{1}$\orcidA{}, Marcelo A. Pires $^{2}$\orcidB{}, Silvio M. Duarte Queir\'os $^{3}$\orcidD{} and Nuno Crokidakis $^{4,}$*\orcidC{}}

\AuthorNames{Firstname Lastname, Firstname Lastname,  and Firstname Lastname}
\AuthorCitation{Lastname, F.; Lastname, F.; Lastname, F.}

\address{%
$^{1}$ \quad Departamento de Economia, Universidade de Bras\'ilia, Bras\'ilia/DF, Brazil; andrelo@id.uff.br\\
$^{2}$ \quad Universidade Federal de Alagoas, Delmiro Gouveia/AL, Brazil; marcelo.pires@delmiro.ufal.br\\
$^{3}$ \quad Centro Brasileiro de Pesquisas F\'isicas, Rio de Janeiro/RJ, Brazil and National Institute of Science and Technology for Complex Systems; sdqueiro@cbpf.br \\
$^{4}$ \quad Instituto de F\'isica, Universidade Federal Fluminense, Niter\'oi/RJ, Brazil; nunocrokidakis@id.uff.br}

\corres{Correspondence: nunocrokidakis@id.uff.br}

\abstract{
We study the Galam's majority-rule model in the presence of an independent behavior that can be driven intrinsically or can be mediated by information regarding the collective opinion of the whole population. 
We first apply the mean-field approach where we obtained an explicit time-dependent solution for the order parameter of the model. 
We complement our results
 with Monte Carlo simulations where our findings indicate  that independent opinion leads to order-disorder continuous nonequilibrium phase transitions. 
Finite-size scaling analysis show that the model belongs to the mean-field Ising model universality class. Moreover, results from an approach with the Kramers-Moyal coefficients provide insights about the social volatility.
}
 
\vspace{1cm} 

\keyword{Social dynamics; Collective phenomena; Phase transitions; Monte Carlo simulation}

\begin{document}


\tableofcontents

\section{Introduction}

Opinion dynamics is one of the hottest topics in Sociophysics. This recent research area uses tools and concepts of statistical physics to describe  some aspects of social and political behavior~\cite{galam2008sociophysics,sen2014sociophysics,castellano2009statistical,DALUZ2023113379}. From the theoretical point of view, opinion models are interesting to physicists because they can present order-disorder transitions, hysteresis, scaling, and universality, among other typical features of physical systems, which have attracted the attention of many groups throughout the world  \cite{doi:10.1142/S0219525921500065,sobkowicz2020whither,martins2020discrete,baumann2020modeling,anagnostopoulos2022biased,zubillaga2022three,2019oestereichPC,oestereich2020hysteresis,vazquez2020role,oestereich2022impact}. Concerning sociologists, these methods are useful to improve forecasting by means of controlled toy-models that can be run multiple times and help fine tune field studies as well~\cite{imai2017quantsoc}. In addition to the interesting properties of opinion dynamics models, per se, such dynamics have also been applied in various fields such finance and business~\cite{zha2020opinion}, epidemic dynamics with the presence of conflicting opinions \cite{fang2023coevolution,2018piresOC,2021piresOCQ,oestereich2023optimal,franceschi2023modeling,harari2022epidemic,wang2021information}, among others~\cite{castellano2009statistical}.

Among the most studied models, we can highlight the voter model~\cite{clifford1973model,holley1975ergodic}, the Sznajd model~\cite{SZNAJDWERON2021125537}, the Deffuant model~\cite{2000deffuantNAW}, the kinetic exchange opinion models \cite{biswas2023social} and the majority rule model \cite{2008galam,GALAM1999132,galam2002minority,PhysRevLett.90.238701}. All the mentioned models are build based on distinct microscopic rules that control the dynamics of interactions among agents. The Sznajd model considers a two-state (up/down spins) outflow dynamics, where a group of agents sharing a common opinion influence the groups neighbors to follow group's opinion. The model presents a phase transition between the positive and the negative consensus: initial densities of spins up smaller than $1/2$ lead eventually to all spins down, and densities greater than $1/2$ to all spins up, i.e., consensus absorbing states where the system cannot escape \cite{SZNAJDWERON2021125537}. On the other hand, the Deffuant model considers the opinions as continuous variables, and the interactions depend on the "distance" among pairs of opinions, which defines the concept of bounded confidence. Depending on the value of such bounded confidence, the population can evolve to consensus (all equal opinions) or to polarization (population divided in two distinct opinions). No phase transition is observed \cite{2000deffuantNAW}. The majority rule model considers groups of $g$ agents, that interaction through a simple rule: all agents in the group follow the local majority. In case of even values of $g$, a probability $k$ defines which opinion will win the debate inside the group. The results of the model, regarding consensus and phase transitions, are similar to the observed in the Sznajd model \cite{GALAM1999132}. Some application of the majority rule model are mentioned in the following. Finally, the kinetic exchange opinion models are based on dynamics of wealth exchange. Interactions are pairwise and considers continuous opinions originally, or discrete three-state opinions ($+1, -1$ or $0$ states) \cite{biswas2023social}. Both formulations lead the population to undergoes order-disorder phase transitions, similar to which occurs in spin models. Absorbing states, where all agents are in the neutral state (all opinions $0$ in the population) are observed. Observe that such absorbing states are distinct to the ones observed in the previous models, where all agents share opinion $+1$ or $-1$. Such kind of consensus states are observed in kinetic exchange opinion models only in very specific situations \cite{biswas2023social}.

We are especially interested in the majority rule model, proposed by Serge Galam \cite{2008galam}. In this model, random groups of agents are chosen and after the interaction of such agents all of them assume the initial majority opinion. The model was studied by many groups~\cite{crokidakis2015inflexibility,2014crokidakisO,abilhoa2020density,galam2012dictatorship,gimenez2021opinion,krapivsky2021divergence,muslim2021phase,doi:10.1142/S0129183123500882}, and it was applied to a series of practical problems, like antivax movement~\cite{pires2017dynamics}, USA~\cite{galam2017trump} and French \cite{galam2018unavowed} presidential elections, terrorism~\cite{GALAM2023113119}, among many others.

Independence in opinion making and the failure of group influence was considered in several opinion dynamics models~\cite{sznajd2011phase,nyczka2013anticonformity,nyczka2012phase,muslim2022opinion,vieira2016phase,muslim2020phase,crokidakis2014phase,abramiuk2020generalized,2016vieiraAC}. 
A recent extension of Galam's model in Ref.~\cite{crokidakis2015inflexibility} considered the impact of independence in social dynamics. In that case, with probability $q$, an individual acts independently of the majority opinion of their group and chooses at random one of the two possible opinions. The introduction of that condition, quantified by the parameter paves the way to the occurrence of an order-disorder nonequilibrium phase transition that does not occur in the original majority-rule model~\cite{2008galam}.

In this work, we go farther afield than the independence mechanism considered in Ref.~\cite{crokidakis2015inflexibility}, and we take into account the overall global opinion of the population when an agent decides to act independently of the group's opinion.
With this we can paint a more detailed picture of the process of independence, since now agents can take global opinion into account when they ignore in-group majority.
This change manages to incorporate the concept of ``impersonal influence'' \cite{1992mutz} established within political science. The goal of which is to quantify the influence of the anonymous mass of individuals outside her small-world composed of family, (close) friends and acquaintances. That impersonal influence encompasses polls, reader's comments on news on digital media and the individual's general perception by consulting social networks that can have an effect on her decision making process.

The strength of this new effect can be controlled  by a new parameter $g$ that gauges the impact of the global population opinion, which is the macrostate, on the individuals. This impact can be of a contrarian nature, for negative values of $g$ where agents tend to to take opposite opinion from the population or it can be positive and reinforce the predominant opinion, thus helping the building of consensus.

With that we go along the lines of canonical considerations over complex systems for which microscopic and macroscopic features influence one another. We develop an analytical framework in order to understand the results from numerical simulations. All results suggest the occurrence of order-disorder transitions, and the estimates of the critical exponents indicate that the model is in the mean-field Ising model universality class.


\section{Model and methods}

Herein, we analyze a majority-rule model with independence; however, differently to Ref.~\cite{crokidakis2015inflexibility} we assume a density-dependent probability $f_t$ for changing the current opinion independently of the interaction group.

\subsection{Model}

Let us consider a population of $N$ individuals, $i$, with opinions $A$ or $B$, with respect to a given issue, that map into a stochastic variable, $o_i$, such that $o_i (A,t) = + 1$ and $o_i (B,t) = - 1$. Macroscopically, we compute   the density of agents with opinion A,
\begin{equation}\label{eq1}
    \eta _A (t) \equiv \frac{1}{N} \sum _{i = 1} ^{N} \delta_{o_i(t),+1},
\end{equation}
and the density of agents with opinion B,
\begin{equation}
    \eta _B (t) \equiv \frac{1}{N} \sum _{i = 1} ^{N} \delta_{o_i(t),-1} = 1 - \eta _A (t).
\end{equation}

The mean opinion, from which we establish the macroscopic state of the system reads,
\begin{equation}
m(t) \equiv \frac{1}{N} \sum _{i = 1} ^{N} o_i(t) = \eta _A (t) - \eta _B (t).
\label{orderparameter}
\end{equation}

The dynamics of each individual is governed at each time step, $t$, by the following set of rules:
\begin{itemize}
\item An individual with opinion $A$ can change to opinion $B$ through two mechanisms:
\begin{itemize}
\item 
with probability $q$ the individual acts independently of their group. In that case, they change their opinion with probability $f^{AB}(t) = f(1-g \,m(t)) $; 

\item otherwise the individual does not act on their own, then there is a probability $1-q$ that they change their opinion according to a local majority-rule, $A+2B \rightarrow 3B$.
\end{itemize}

\item On the other hand, an individual with opinion $B$ can flip to opinion $A$  through 2 mechanisms:
\begin{itemize}
\item with probability $q$ they decide to whether act independently of their group or not. In that case, the agent will change their opinion with probability $f^{BA}(t) = f(1 + g \,m(t)) $;

\item else if the individual does not act on their own, then there  is a probability $1-q$ that they change their opinion according to a local majority-rule, $B+2A \rightarrow 3A$.
\end{itemize}
\end{itemize}

The rules above are translated into the transition matrix,
\begin{equation}
    W(t) \equiv
    \begin{bmatrix}
    w_1 & w_2 \\
    w_3 & w_4
    \end{bmatrix} =
    \begin{bmatrix}
    q\, f^{AB}(t) & 1-q \\
    1-q & q\, f^{BA}(t)
    \end{bmatrix}. 
\end{equation}

Note that the definitions $f(t)^{AB} = f(1-g \, m(t)) $  and $f(t)^{BA} = f(1+g \, m(t)) $ imply that if 
$$\eta _A(t)> \eta _B(t) \Rightarrow m(t)>0  \Rightarrow f(t)^{BA} > f(t)^{AB} $$ as expected. 

Let us have a closer look at the parameters involved in the model: the parameter $q$ is related to the backbone of our approach establishing the relative weight of the local peer-pressure, $p=1-q$, leading to a decision-making process wherein the individual either submits to the local majority (a conformist behavior) or the decision-making dynamics is carried out on her own. The probability $f^{XY} (t)$ -- related to the latter case -- is naturally shaped by the assessment of the state of affairs provided by the global state, $m(t)$, so that a standard propensity to change opinion through reflection, $f$, is either boosted or mitigated. Epistemologically, the shaping of the probability is equivalent to the process of risk-taking versus risk-aversion described within prospect theory~\cite{kahneman1979prospect}. Herein, we assume a linearized form $f^{XY} (m(t)) = f + \upsilon \, m(t) + \mathcal{O}(m(t)^2)$; depending on the sign of $\upsilon $ we have either a follower or a contrarian impact.
If $g=0$, then $f^{XY} (t) = f$ and we recover the results of~\cite{crokidakis2015inflexibility}.

\subsection{Simulation details}

Our Monte Carlo simulations are structured within an agent-based framework, as individuals constitute the underlying object of study in social theories~\cite{conte2012manifesto}. In our algorithm we consider a computational array of size $N$ to store the opinion of each agent. In each time $t$ we apply a Monte-Carlo step (MCS) that represents a complete iteration through all agents. During each interaction, the simulation chooses a group of $3$ agents at random, considering their current opinion and applying specific rules. These rules are summarized in Table~\ref{tab:model} and define how an agent's opinion may change based on various conditions and probabilities. After each MCS  we implement a simultaneous-parallel updating. This means that the updated opinions are applied to all agents at the same time, ensuring that the changes in opinions are synchronized across the entire population.


\begin{table}[h]
\centering
\caption{red}{Agent-based rules of our model}

\begin{tabular}{p{0.25\textwidth} p{0.45\textwidth} } \hline  
  \multicolumn{2}{|l|}{Each agent with opinion $A$ can flip to opinion $B$ through two mechanisms:} \\ \hline
   \textbf{1.} $A \rightarrow B$ &  $ p^{(1)}_{A \rightarrow B}  = q \, f^{AB}(t) $  \\ \hline
   \textbf{2.} $A+2B \rightarrow 3B$ & $ p^{(2)}_{A \rightarrow B} = (1-q)  \, \eta _B (t)^2 $ \\
  \hline
  \multicolumn{2}{|l|}{Each agent with opinion $B$ can flip to opinion $A$ through two mechanisms:} \\ \hline
   \textbf{1.} $B \rightarrow A$ & $ p^{(3)}_{B \rightarrow A} = q \, f^{BA} (t) $   \\
  \hline
   \textbf{2.} $B+2A \rightarrow 3A$ & $ p^{(4)}_{B \rightarrow A} = (1-q) \,  \eta _A (t)^2 $  \\
  \hline
\end{tabular}

\label{tab:model}
\end{table}

\vspace{1.0cm}


\section{Results and discussion}

\subsection{Analytical results}

Using the Mean-Field approach we can obtain a set of ordinary differential equations that describes the time evolution of the competing opinions in a population.
To derive the  rate of change of opinions $A$ and $B$ at time $t$ we need to consider  that each opinion is influenced by: 
the intrinsic independent behavior (controlled by the parameter $f$), information-driven independence (modulated by the parameter $g$) and local interactions. Thus, based on the rules summarized in Table~\ref{tab:model} we obtain the following Mean-Field equations:

\begin{align}
 \frac{d \eta _A (t)}{dt}  
&= q \, f^{BA} (t)  \, \eta _B (t) + (1-q) \,  \eta _A (t)^2 \, \eta _B (t)
-  q \, f^{AB}(t)  \, \eta _A (t) - (1-q) \, \eta _A (t) \, \eta _B (t)^2 ,
\\
 \frac{d \eta _B (t)}{dt}  
&= q \, f^{AB}(t) \, \eta _A (t) + (1-q) \, \eta _B (t)^2 \, \eta _A (t) 
-  q \, f^{BA}(t) \,  \eta _B (t) - (1-q) \, \eta _B (t) \, \eta _A (t)^2 ,
\\
f^{AB}(t) &= f(1 - g \, m(t)) ,
\\
f^{BA}(t) &= f(1 + g \, m(t)) .
\end{align}

From Eqs. (\ref{eq1}) - (\ref{orderparameter}), namely that
\begin{equation}
  \eta _A (t) =\frac{1}{2}(1+m(t)), \quad \eta _B (t) =\frac{1}{2}(1-m(t)), \quad \eta _A (t) \, \eta _B (t) =\frac{1}{4}(1 - m(t))^2,
\end{equation}
the set of differential equations yields the ordinary differential equation for the macroscopic variable, $m(t)$. 

\begin{align}
\frac{d m(t) }{dt}  &=   
-2 \, q \, f \, (1-g) m(t) 
+ 
(1-q) \, m(t) \frac{1-m(t)^2}{2}.
\end{align}
In other words, starting from a given condition, $m(0) = m_0$, the macroscopic state evolves and eventually reaches a stationary state $dm/dt = 0$; that state is lower bounded by the maximal state of disagreement then $m = 0$, whereas when the population presents unanimity, $|m|=1$. Thus, we expect that for certain conditions dictated by the parameters of the problem, the system can evade the final stationary state of disagreement and end up in a situation for which $|m| \neq 0$, i.e., a majority of individuals favoring A(B). Physically, $m(t)$ is thus defined as an order parameter.
That turns out clearer when we consider that the population adjust is macrostate $m$ aiming at minimizing its so-called Hamiltonian function. 

That is best understood when we recast the previous equation into
\begin{align}
\frac{d m(t) }{dt}  &= - \frac{\partial \mathcal{H}}{\partial m}  = r \, m(t) + u \, m(t) ^3
\label{eq:dynamics}
\end{align}
where
\begin{equation}
    r = \frac{1}{2} \{q [ 4 f (1-g)+1 ]-1\}, \qquad u = \frac{1}{2} (q-1).
    \label{eqs:r_u}
\end{equation}

\begin{figure}
    \centering
    \includegraphics[width=1.1\textwidth]{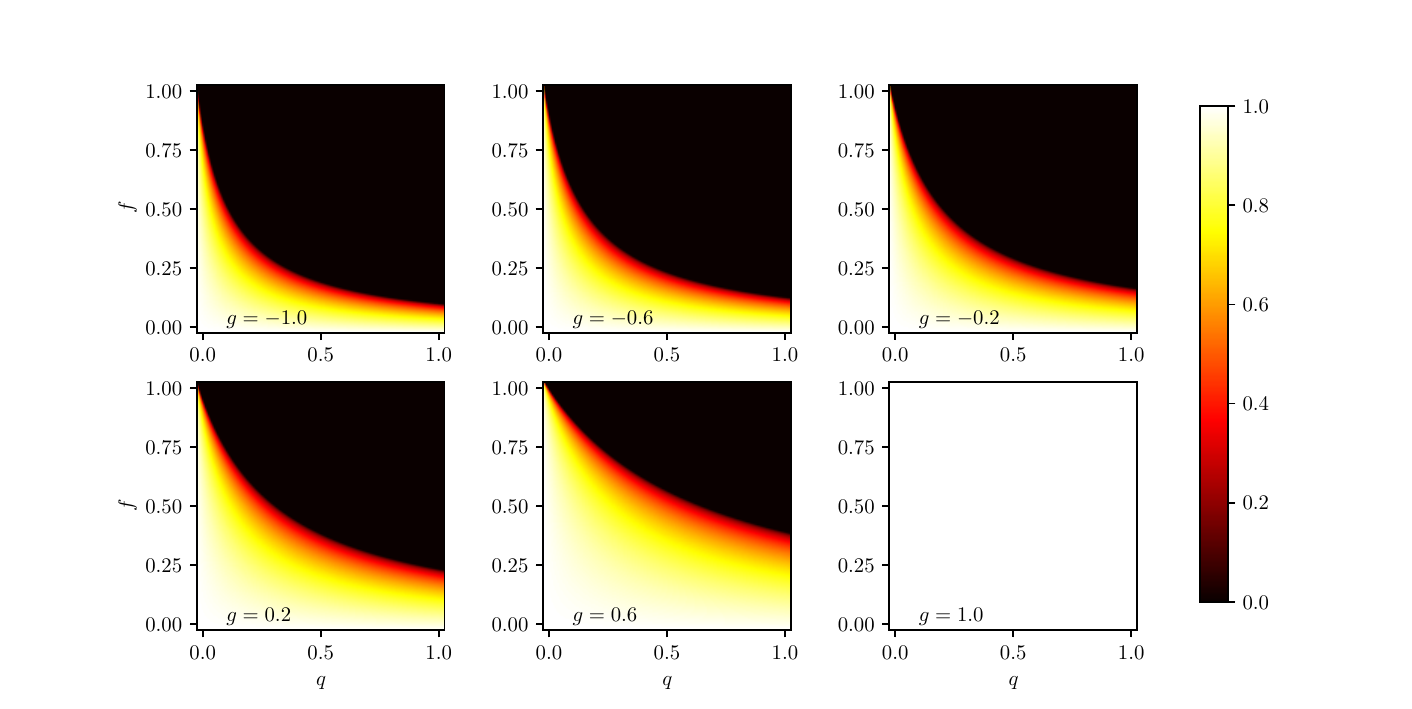}
    \caption{Stationary state solution $|m|$ in the plane $f$ vs $q$, for typical values of $g$. The panels are graphical representations of \Cref{eq:crit_sol}. As negative values of $g$ imply a contrarian effect and positive a follower we notice an increase in the ordered region with an increase in $g$.}
    \label{fig:statstate}
\end{figure}

Therefore, the analytical form of $\mathcal{H}$,
\begin{equation}
    \mathcal{H}(m) = - \frac{1}{2}r \, m^2 - \frac{1}{4} u \, m^4, 
\end{equation}
dictates not only the dynamics of the parameter $m$, but its stable outcome.
First, since $q \le 1 $ and $u < 0$, the stability of the process is assured as the fourth-order term is positive. In the limit $q \rightarrow 1$, the agents will act independently from the local group -- and totally rely on their assessment of the position of the whole population -- we have $\lim _{q \rightarrow 1} u = 0^-$ and $\lim _{q \rightarrow 1} r >0$, which opens the door to non-trivial minima of $\mathcal{H}$ at $ m _c \neq 0$. In being $u < 0$, the emergence of those $m \neq 0$ minima are related to the change of convexity of $\mathcal H$ at $m = 0$ from $\frac{d^2\mathcal H}{dm^2} \large\vert _{m=0} >0$ to a concave profile $\frac{d^2\mathcal H}{dm^2} \large\vert _{m=0} < 0$. The fulfillment of the concave condition implies 
\begin{equation}
    \vert m \vert = m_c=\sqrt{- \frac{r}{u}} = \sqrt{1 - \frac{4fq(1-g)}{1-q}} , \qquad \vert m \vert \le 1.
\label{eq:crit_sol}
\end{equation}
The graphical representation of which can be seen in Figure \ref{fig:statstate}. After plugging the relations in Eq.~(\ref{eqs:r_u}) it reads

\begin{align} 
m \sim
\left( q_{c}-q \right)^{\beta}
\label{Eq:order_par}
\end{align}
where  $\beta=1/2$ and 
\begin{equation} 
q_{c} = \frac{1}{1+4f(1-g)},
\label{Eq.limiar}
\end{equation}
which defines the critical peer-pressure relative weight, $p_c \equiv 1-q_c$.
Heed that instances where $g < 0$, imply in a smaller value of $q_c$ and therefore a larger $p_c$ in what we assume as a freethinker-prone behavior; on the other hand, when $g > 0$ we regard it as a conformist-prone case. 

Eq. (\ref{Eq:order_par}) with $\beta=1/2$ suggests a phase transition in the same universality class of the mean-field Ising model. We will discuss this point in more details in the following, when we will exhibit the results of Monte Carlo simulations of the model.

Equation~(\ref{Eq.limiar}) corresponds to the limit $t \rightarrow \infty $ of the solution to Eq.~(\ref{eq:time_sol}) which reads,
\begin{equation}
    m(t) = \left[\exp(-2 r t) \left(\frac{1}{m_0^2}+\frac{u}{r}\right)-\frac{u}{r}\right]^{-1/2} = m_c \left[ \exp(-2rt) \, \left( \frac{m_c^2}{m_0^2} -1 \right) + 1 \right] ^{-1/2},
    \label{eq:time_sol}
\end{equation}
where $m_0$ is the macroscopic initial condition of the system.

We can further explore the dynamical behavior of the system, especially when the parameters are set at their critical values and one lets the system evolve. In that case, two situations deserve particular attention: when the initial state corresponds to unanimity, $m_0 = 1$, the factor given by $\exp[-2rt] \, \left( \frac{m_c^2}{m_0^2} -1 \right)$ in Eq.~(\ref{eq:time_sol}) can be seen as perturbation, whereas for the same factor dominates Eq.~(\ref{eq:time_sol}) when the initial condition is that of full disagreement ($m_0 \rightarrow 0$). That results in two quite different behavior of $m(t)$ in the short-term.

\subsection{Probabilistic approach}

The previous deterministic approach can be further seasoned when fluctuations are taken into account. Recalling  for a population of $N$ individuals the macroscopic state, $m$, changes by $ \mu =\pm 2/N$ every time an individual switches their opinion with each opinion fraction varying by $1/N$, if we focusing on the time evolution of the fraction of individuals with opinion $A$ at time $t+1$ it reads,
\begin{equation}
\eta _A (t+1) - \eta _A(t) = \frac{1}{N} p^\dagger (t) - \frac{1}{N} p (t),
\end{equation}
where
\begin{equation}
p^\dagger (m,t) = w_1 \, \eta _A + w_2 \, \eta _A \, \eta _B ^2,
\end{equation}
corresponds to the probability that the number of people with opinion $A$ increases by one individual whereas
\begin{equation}
p (m,t) = w_3 \, \eta _B \, \eta _A ^2 + w_4 \, \eta _B ,
\end{equation}
gives one the probability that the number of people with opinion $A$ diminishes by one individual. These quantities are identified as operators of creation and destruction in the probability space~\cite{vankampen}.

Taking into consideration that $p^\dagger $ and $p $ correspond to an increment and a reduction of the macroscopic state by $\mu = 2/N$, respectively, we can establish the following master equation for the evolution of $m$ for a time step $\epsilon = 1/N$,
\begin{equation}
\eta (m, t+\epsilon) =
 p^\dagger(m-\mu, t) \, \eta (m-\mu,t) + p (m+\mu, t) \, \eta (m+\mu,t) + \bar{p} (m,t) \eta (m,t),
 \label{mastereq}
\end{equation}
with $\bar{p} \equiv 1 - p^\dagger  - p$ quantifying to the maintenance of the macroscopic state. Formally, Eq.~(\ref{mastereq}) fits within the (normalized) one-step class of stochastic processes and thus,
\begin{equation}
    \eta (m,t) = \exp \left[ \mathbf{L}_{KM} (m,t)\right] \eta (m_0 ,0) \qquad \qquad \eta (m,0) = \delta(m - m_0).
\end{equation}
where, bearing in mind we are computing a normalized quantity and not simply $N_A-N_B$, the Kramers-Moyal operator reads
\begin{equation}
    \mathbf{L}_{KM} (m,t) = \sum _ {n=1} ^\infty \frac{(-\mu)^{-n}}{n!} \frac{\partial^n}{\partial m^n} \left[ p_{m,t} + (-1)^n  \,  p^\dagger_{m,t}  \right].
    \label{fpe-formal-sol}
\end{equation}

In considering $\mu \rightarrow 0$ so that the variance of $m_t$ is kept fixed and equal to $\sigma _m ^2 (t)$, we neglect the terms of order $n>2$ and the formal solution gets the form of a Fokker-Planck Equation,
\begin{equation}
\frac{\partial \eta (m,t)}{\partial t} = -\mu ^{-1} \frac{\partial }{\partial m} \left[ D_1(m,t) \, \eta (m,t) \right] + \frac{\mu ^{-2}}{2} \frac{\partial ^2}{\partial m^2} \left[ D_2(m,t) \, \eta (m,t) \right].
\end{equation}

Therefrom we identify,
\begin{equation}
D_1(m,t) \propto p(m,t)  - \,  p^\dagger(m,t)
\label{D1}
\end{equation}
that defines the shape of the effective potential wherein the macroscopic dynamics of the order parameter evolves in time; on the other hand,
\begin{equation}
D_2(m,t) \propto p(m,t)  + \,  p^\dagger(m,t)
\label{D2}
\end{equation}
characterizes the magnitude of the fluctuations, which in the present social system we associate to the concept of social volatility~\cite{behrens2008volatility}.

Plugging the previous relations for the probability creation/annihilation operators into Eqs.~(\ref{D1})-(\ref{D2}) we finally get,
\begin{equation}
D_1(m,t) \propto r \, m(t) +u \, m(t)^3
\end{equation}
as given by the effective Hamiltonian Landau approach. Regarding the second order term,
\begin{equation}
D_2(m,t) \propto 1 + q\, (4f-1) - [1+q(4\,f\,g-1) ] \, m(t)^2 
\label{multiplicative}
\end{equation}
Equation~(\ref{multiplicative}) indicates a macroscopic feature of this model that is worth noting: the magnitude of the fluctuations -- i.e., the social volatility -- exhibited by the system depends on its state in such a form that as $m$ increases and approaches $m=1$ (unanimity), the volatility decreases. On the contrary, when the group shows strong disagreement, \mbox{$m \approx 0 $}, they approach the sate of maximal volatility. That behavior contrasts with what is measured in quantitative finance where the realized volatility is directly proportional to price variations~\cite{queiros2005quantfin}.
If we heed that within a physical context $D_2$ is related to the (local) temperature of a physical system, we assert that our model is able to capture the cooling down and the heating up of a social system as it approaches or departs from consensus.

Alternatively, the fluctuations given by $D_2(m,t)$ can be understood from another perspective: bearing in mind that $p^\dagger $ and $p$ respectively correspond to an increment and a reduction of the macroscopic state by $\mu = 2 /N $, we then interpret $D_1$ as the imbalance between the likelihood of increment and reduction of $p(m)$ whereas $D_2$ is related to the average over increment and reduction, which is clearly non-vanishing. That is related to the microscopic change of opinion that each individual can make and which corresponds to a source of the macroscopic fluctuations that end up being expressed by the social volatility.
Complementary, those fluctuations yield an entropy production that can be associated with the total information output due to the microscopic interaction between agents. Therefore, around a consensus we measure a less volatile state as it is less entropic and vice-versa.

\subsection{Monte Carlo simulation and finite-size scaling}

\begin{figure}[!htb] \centering
\includegraphics[width=0.4\textwidth]{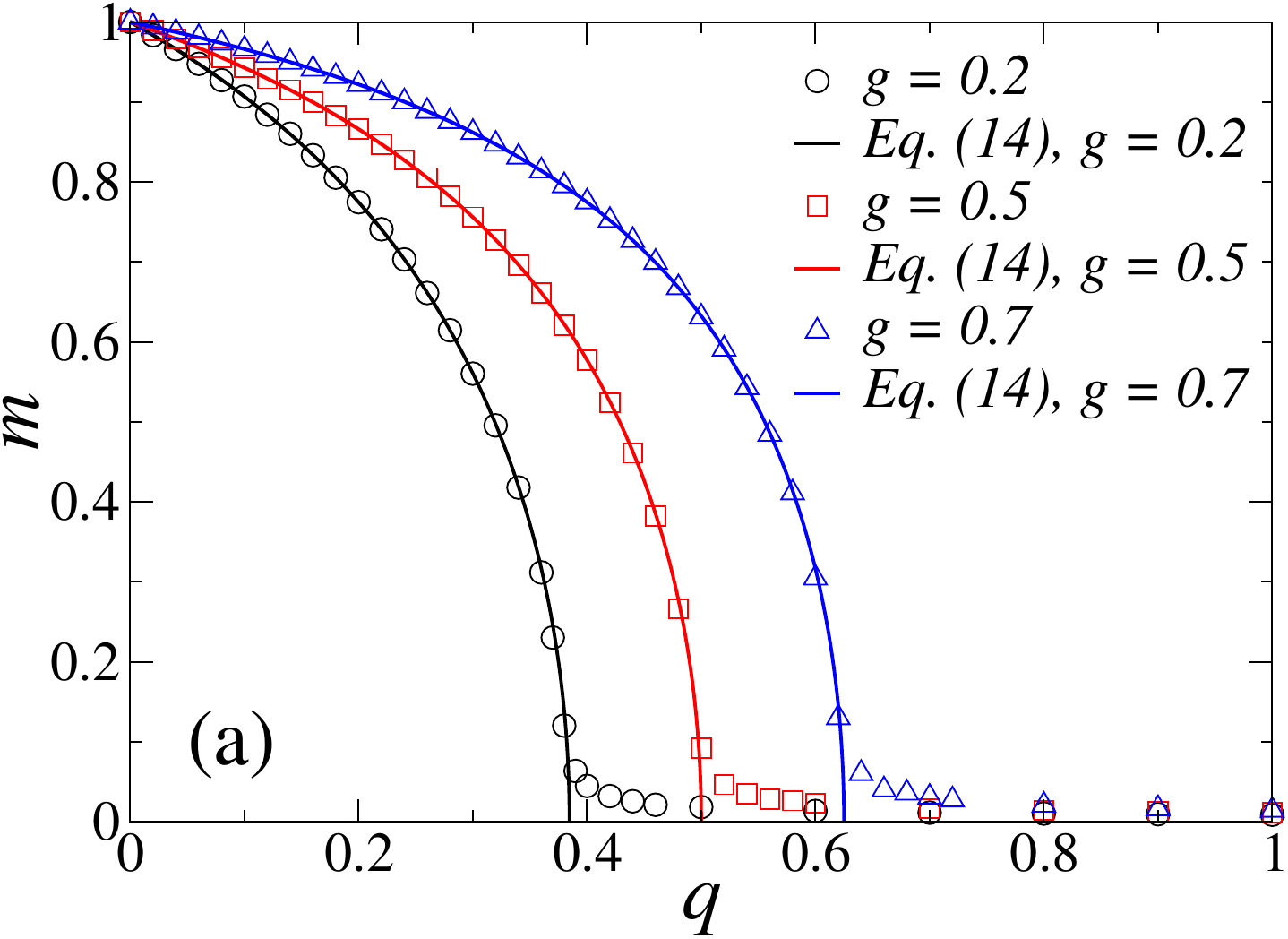}
\includegraphics[width=0.4\textwidth]{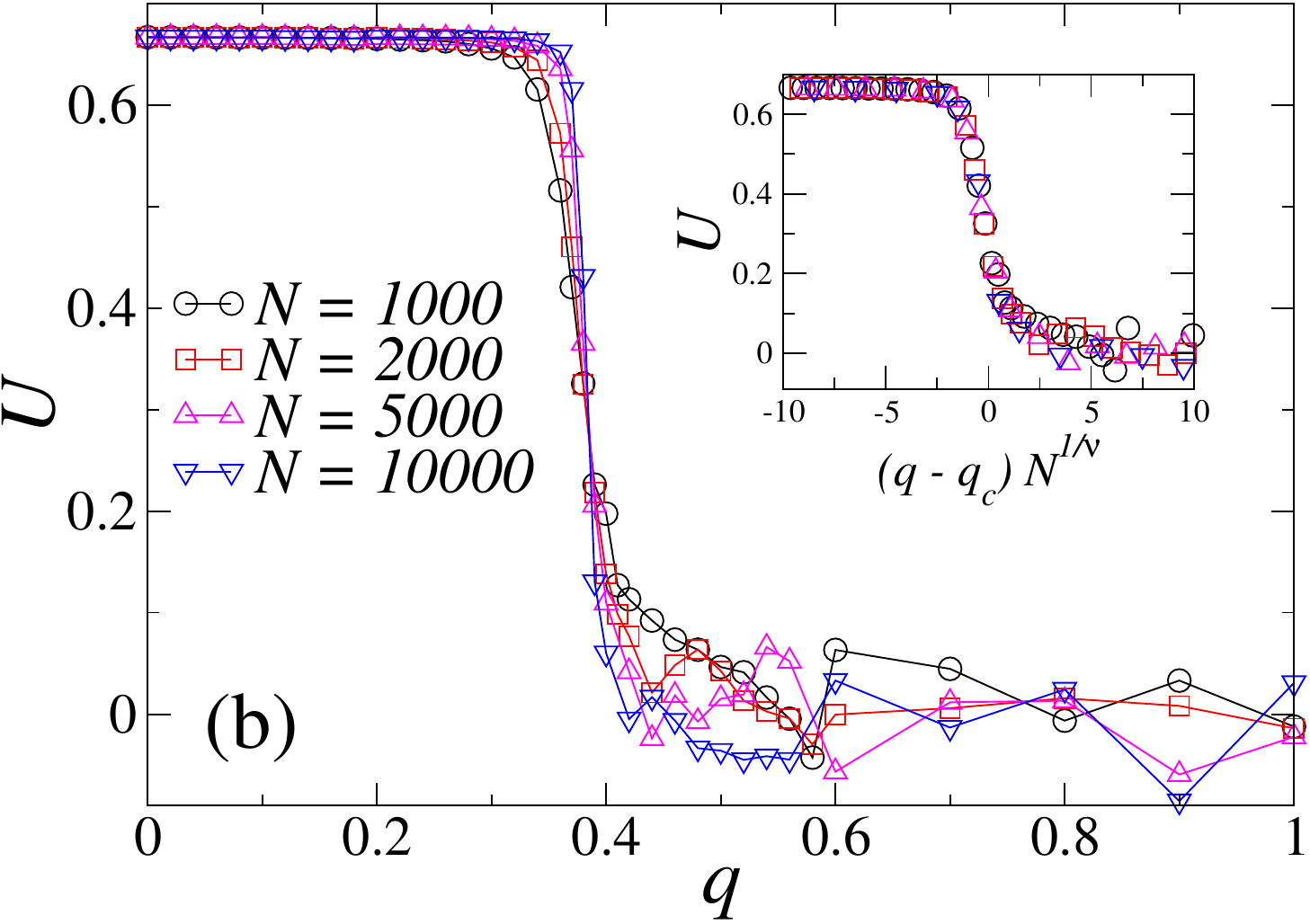}
\\
\includegraphics[width=0.4\textwidth]{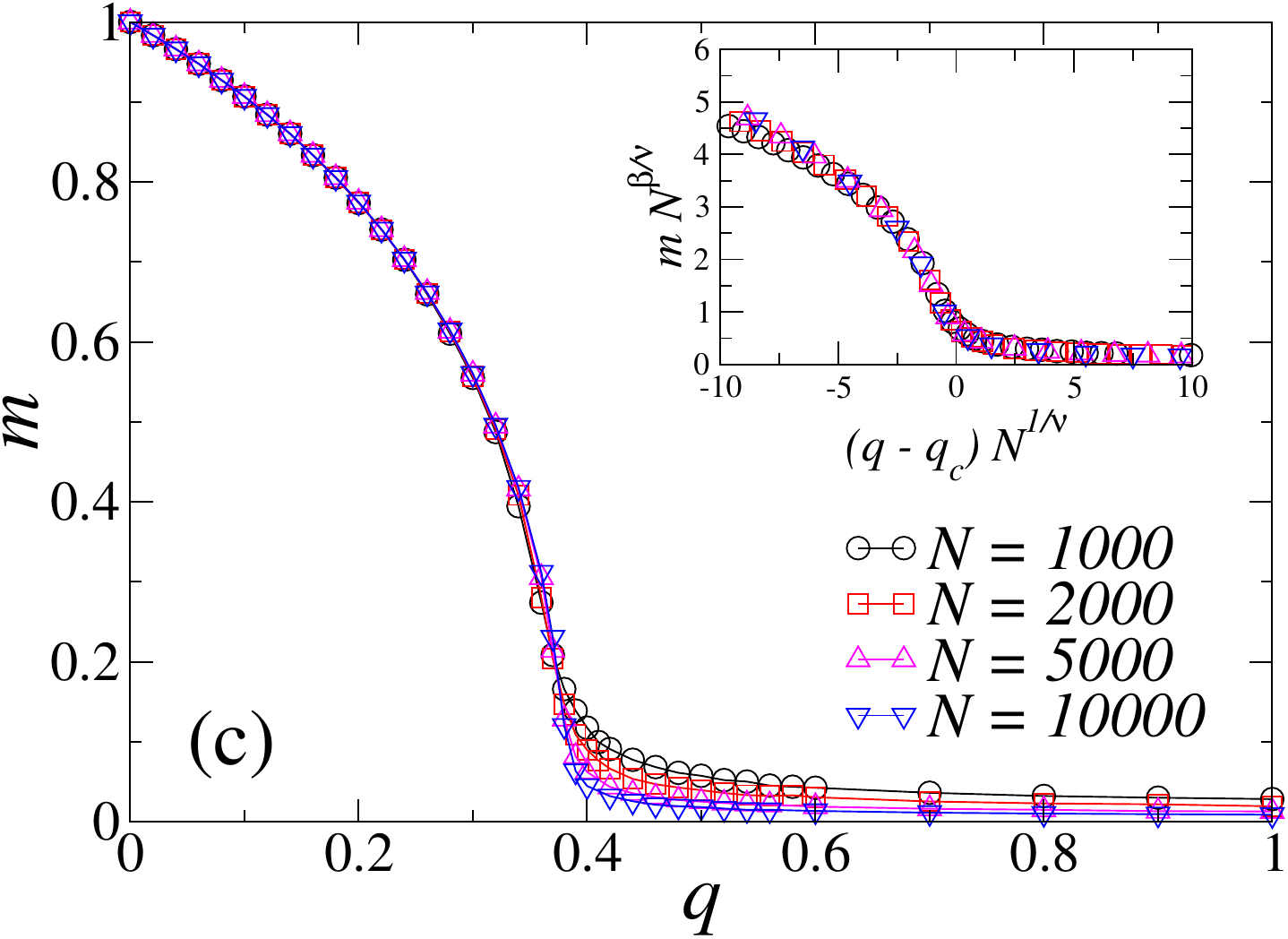}
\includegraphics[width=0.4\textwidth]{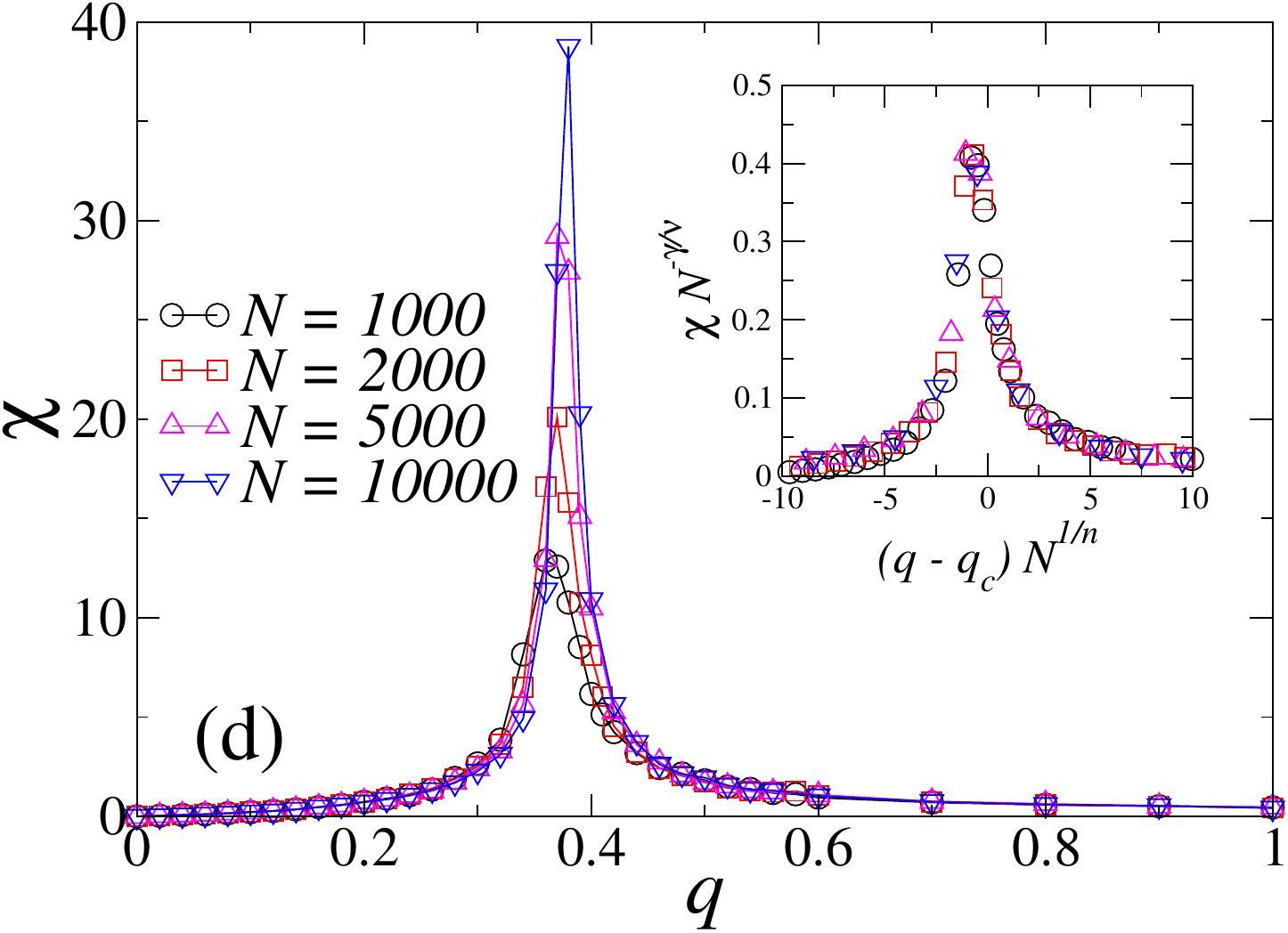}
\caption{Results of Monte Carlo simulations of the model for $f=0.5$. (a) Stationary order parameter $m$ (collective opinion) as a function of $q$. The symbols come from simulations for $N=10^{4}$ and typical values of $g$, and the lines are obtained from the analytical result, Eq. (\ref{eq:crit_sol}). We also show the results for fixed $g=0.2$ and distinct population sizes $N$, and the corresponding finite-size scaling analysis for the Binder cumulant $U$ (panel (b)), the order parameter $m$ (panel (c)) and the susceptibility $\chi$ (panel (d)). We obtained $q_c \approx 0.385$, $\beta \approx 0.50$, $\gamma \approx 1.00$ and $\nu\approx 2.00$. Data are averaged over 100 simulations.}
\label{fig:m-vs-q}
\end{figure}

In Fig.\ref{fig:m-vs-q} (a) we provide a comparison between the analytical solution, as elucidated in the preceding section, and the Monte Carlo simulation for Galam's model with information-mediated independence. We plot the stationary values of the macrostate obtained from simulations for a population size $N=10^{4}$ and from Eq. (\ref{eq:crit_sol}) for typical values of $g$ and fixed $f=0.5$. We see a good agreement between the results obtained in both ways. We observe clear order-disorder phase transitions as well, which denotes a collective change in the population behavior. These transitions denote a macroscopic change from the so-called ordered state characterized by the presence of a well-defined majority ($|m|>0$) to the disordered-state characterized by the absence of a clear majority ($|m|\sim 0$). When $q=0$, we recover the usual result obtained from Galam's model., i.e., a consensus in the population (all agents sharing opinion A or B).

In order to verify the universality class of the model, we have performed numerical simulations for distinct population sizes and applied a so-called scaling analysis. In addition to the order parameter, $m$, we have also computed the fluctuations $\chi$ of the order parameter (or ``susceptibility''), defined as
\begin{equation}
\label{eq:suscept}
\chi \equiv  N\,(\langle m^{2}\rangle - \langle m \rangle^{2})   
\end{equation}
and the Binder cumulant $U$, defined as \cite{binder1981finite}
\begin{equation} \label{eq:binder}
U   \equiv   1 - \frac{\langle m^{4}\rangle}{3\,\langle m^{2}\rangle^{2}} \,.
\end{equation}

As an example, we exhibit in Fig.~\ref{fig:m-vs-q} the finite-size scaling (FSS) analysis of the order parameter, the susceptibility and the Binder cumulant for four lattice sizes, for $f=0.5$ and $g=0.2$. We have identified the critical value $q_{c}$ by the crossing of the Binder cumulant curves, as can be seen in the main panel of Fig. \ref{fig:m-vs-q} (b). We have obtained $q_c\approx 0.385$, in excellent agreement with the analytical result of Eq. (\ref{Eq.limiar}), that gives us $q_c\approx 0.3846$. The critical exponents $\beta$, $\gamma$ and $\nu$ were found by the best collapse of data. The FSS analysis was based on the standard relations,
\begin{eqnarray}
m(q,N) & \sim & N^{-\beta/\nu} \\
\chi(q,N) & \sim & N^{\gamma/\nu} \\
U(q,N) & \sim & constant \\
q_c(N) - q_c & \sim & N^{-1/\nu} \label{corlength}
\end{eqnarray}
Considering the above equations, we obtained $\beta\approx 0.5$, $\gamma\approx 1.0$ and $\nu\approx 2.0$. The data collapses are exhibited in Fig.  \ref{fig:m-vs-q}, panels (b) - (d). We also verified that for other values of $f$ and $g$ the same exponents are obtained. The results suggest that the model belongs to the mean-field Ising model universality class, as well as it is in the same universality class of the Sznajd model and kinetic exchange opinion models in the presence of independence \cite{CALVELLI2019518,sznajd2011phase,muslim2022opinion,crokidakis2014phase}.

The above results, namely Eq.~(\ref{corlength}) bridge with the dynamical analysis as Eq.~(\ref{eq:time_sol}) sets up a relaxation time scale, $\tau $ of the macroscopic parameter that is inversely proportional to $r$. 
Comparing the exponential factor in Eq.~\ref{eq:time_sol} with the usual term related to the relaxation $e^{t/\tau}$ we obtain the relaxation time $\tau=-1/(2r)$. Then, plugging Eq.~(\ref{Eq.limiar}) into the definition of $r$ we get $r=(q-q_c)/(2q_c)$ and finally 
\begin{equation}
    \tau \sim (q_c-q) ^{-1}.
\end{equation}
Explicitly, at criticality we have a relaxation time scale of the order parameter $m(t)$  that diverges with the same scale-invariant functional form as the correlation length. Because the propagator given by the Fokker-Planck Equation rules all relaxation quantities of $m(t)$, the same slowing down near the transition is found for the self-correlation function of m(t), $\left\langle m(t^\prime) \, m(t) \right\rangle \sim \exp [ - |t^\prime - t| / \tau]$.

\FloatBarrier


\section{Conclusions}

In this work we have studied an extension of the Galam's majority-rule model. For this purpose, we introduced the mechanism of independence, considering that individuals can act independently of their interaction groups with a given probability $q$ that is complementary to the peer-pressure weight, $p=1-q$. In addition, the individual inspects the global population opinion and such opinion affects their independent probability. When an individual does not act independently of the group, she follows the local majority opinion, as in the original Galam model.

We have observed that the independence mechanism leads the population to undergo a critical change of behavior at $q=q_c$ in which a minimal consensus $m \neq 0$ -- where $m$ is the order parameter of the model --, optimizes the overall state of the population best than the case of complete disagreement. Within that phase transition context, we have derived an expression for order parameter $m(t)$. From its stationary solution, we have obtained the critical behavior $m \sim (q_c - q)^{\beta}$ with $\beta = 1/2$. We understood that as one approached the critical transition, the relaxation of the overall state is ever slower with its typical time scale $\tau \sim (q_c - q)^{-1}$. The other canonical critical exponents $\gamma$ and $\nu$ were obtained through Monte Carlo simulations. From the set of critical exponents, we have verified that the model is in the ubiquitous universality class of the mean-field Ising model. That result is expected since as opinions are mapped into random variables, $o_i = \pm 1$, the phase transition corresponds to a group $\mathbb{Z}_2$ symmetry breaking of which the Ising model is the quintessential case.

We mention that while our model is not defined by a physical Hamiltonian, the identification with the Ising universality class arises from a series of results coming from 3 methods: mean-field approach, Monte Carlo simulations, and finite-size scaling analysis.

From the microscopic dynamics, we have derived the probabilistic evolution of $m$. Those results allowed us to confirm the critical behavior of $m$ from the first Kramers-Moyal coefficient and from the second the nature of the fluctuations that can be coined as social volatility. In respect of the latter, we have learned that the magnitude of the volatility depends on the state of the population in a inverse proportion relation way, so that in this case herding in opinion tends to induce less agitation in the population.
Further insights into this subject-matter will be discussed in future work.

\reftitle{References}

\bibliography{bibliography}

\end{document}